\documentclass[conference]{IEEEtran}
\ifCLASSINFOpdf
\else
\fi
\usepackage{makeidx}         
\usepackage{graphicx}        
\usepackage{multicol}        
\usepackage[bottom]{footmisc}
\usepackage{algorithm,algorithmicx,algpseudocode}
\usepackage{graphicx}
\usepackage{subfigure}
\usepackage{amssymb}
\usepackage{multirow}
\usepackage{hyperref}
\usepackage{amsmath}
\usepackage{adjustbox}

\usepackage{tabularx}

\usepackage{array}
\newcommand\MyBox[2]{
  \fbox{\lower0.75cm
    \vbox to 1.7cm{\vfil
      \hbox to 1.7cm{\hfil\parbox{1.4cm}{#1\\#2}\hfil}
      \vfil}%
  }%
}

\begin{document}
%
\title{A Short Survey on Data Clustering Algorithms}

\author{\IEEEauthorblockN{Ka-Chun Wong}
\IEEEauthorblockA{Department of Computer Science\\
City University of Hong Kong\\
Kowloon Tong, Hong Kong\\
Email: kc.w@cityu.edu.hk}
}


%


\maketitle

\begin{abstract}

With rapidly increasing data, clustering algorithms are important tools for data analytics in modern research. They have been successfully applied to a wide range of domains; for instance, bioinformatics, speech recognition, and financial analysis. Formally speaking, given a set of data instances, a clustering algorithm is expected to divide the set of data instances into the subsets which maximize the intra-subset similarity and inter-subset dissimilarity, where a similarity measure is defined beforehand. In this work, the state-of-the-arts clustering algorithms are reviewed from design concept to methodology; Different clustering paradigms are discussed. Advanced clustering algorithms are also discussed. After that, the existing clustering evaluation metrics are reviewed. A summary with future insights is provided at the end.

\end{abstract}


%
\IEEEpeerreviewmaketitle

\section{Introduction}
Nowadays, with the support of science and technology, large amounts of data has been, and will continue to be, accumulated. 
For example, a single human genome accounts for about four gigabytes data space \cite{pmid24389653,Wong:2010:ESL:2128650.2128705,Wong:2011:GLP:2093590.2093604} and the transaction logs in financial markets are measured in billions each day  \cite{financialLogs}. Such a large amount of data is overwhelming and prevents us from applying traditional analysis techniques. Scalable methods need to be devised to handle it.
As one of the main analysis tools, cluster analysis methods have been proposed to separate the large amount of data into clusters.
The data clustering methods are unsupervised which means there is not any label for model training; we do not even know the exact number of clusters beforehand.
Given a set of data, a clustering method is expected to divide the data into several clusters by itself. Formally speaking, given a set of data instances, a data clustering method is expected to divide the set of data instances into the subsets which maximize the intra-subset similarity and inter-subset dissimilarity, where a similarity measure is defined beforehand.

\section{Clustering Paradigms}

Since most data clustering problems have been shown to be NP-hard \cite{NPhard}, different methods have been proposed in the past. In general, those methods can be categorized into different paradigms: Partitional Clustering, Hierarchical Clustering, Density-based Clustering, Grid-based Clustering, Correlation Clustering, Spectral Clustering, Gravitational Clustering, Herd Clustering, and Others.

\subsection{Partitional Clustering}

Data is divided into non-overlapping subsets such that each data instance is assigned to exactly one subset. For example, k-means \cite{kmeans} is a classical partitioning method that applies an iterative refinement approach with two main steps. The first step is to choose the means of clusters as the centroids, whereas the second step is to assign data points to their nearest centroids. In practice, its computational speed and simplicity appeal to people \cite{kmeansUse,XiongWC09}.  Its main drawback is the vulnerability to its random seeding technique. In other words, if the initial seeding positions are not chosen correctly, the clustering result quality will be affected adversely. 

In light of that, David Arthur and Sergei Vassilvitskii proposed a method called k-means++ \cite{kmeansPP} to improve k-means in 2007. From section 2.1.and 2.2 in \cite{kmeansPP}, we can observe that the steps 2-4 of k-means++ are exactly the same as those of k-means. The main difference lies in the step 1 which is the seeding technique. A new seeding technique is proposed to replace the arbitrary seeding technique of k-mean. Given a set of seeds chosen, the seeding technique favors the data points which are far from the seeds already chosen. Thus the seeds are chosen probabilistically as dispersed as possible.

\begin{figure}[h!]
\centering
\includegraphics[width=0.48\textwidth]{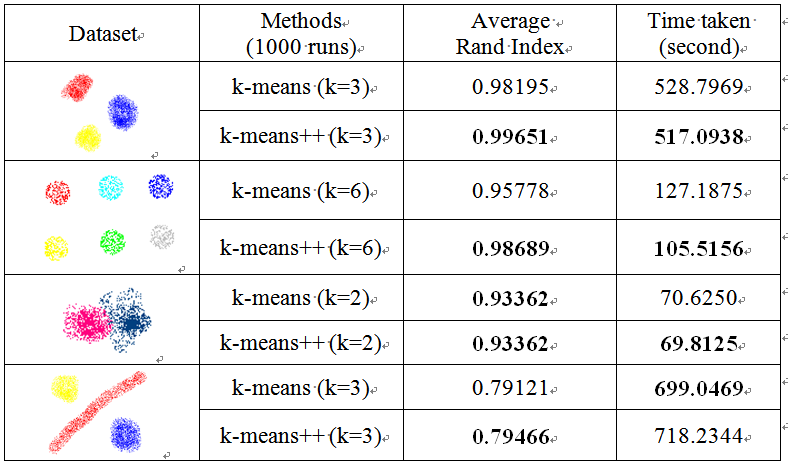}
\caption{Performance Comparison between k-means and k-means++.}
\label{fig:kmeansComparison}
\end{figure}

As k-means++ is the extended version of k-means method, we conducted numerical experiments to evaluate and compare their performance under 1000 replicate runs. For better visual inspection and visualization, the datasets and performance values are both depicted and tabulated in Fig. \ref{fig:kmeansComparison}. We can observe that k-means++ does perform better than k-means on the first three datasets. Both the clustering score (Rand Index) and time taken have been improved. However, the performance comparison is relatively complicated on the last dataset.

\subsection{Hierarchical Clustering}

Clusters are formed by following either a bottom-up approach or a top-down approach. For example, single-linkage clustering  \cite{clusteringReview} is a classic bottom-up approach in which data points are gradually agglomerated together to form clusters. In each step, all pair-wise distances are computed to identify the minimum. The parties involved in the minimal pair-wise distance are linked together. Such a step is repeated until all data points are linked together. A hierarchical tree is constructed to connect all data points at the end. A tree depth level can be chosen to cut the tree, forming clusters. To model data dynamically, a special hierarchical clustering method called Chameleon has been proposed \cite{Chameleon}. It makes use of the inter-connectivity and closeness concept to merge and divide clusters. If the inter-connectivity and closeness between two clusters are higher than those within the clusters, then the two clusters are merged.

\subsection{Density-based Clustering}

Apart from the well-known clustering methods, there are different clustering paradigms. In density-based clustering, data is clustered based on some connectivity and density functions. For example, DBscan \cite{DBscan} uses density-based notions to define clusters. Two connectivity functions \emph{density-reachable} and \emph{density-connected} have been proposed to define each data point as either a core point or a border point. DBscan visits points arbitrarily until all points have been visited. 
If the point is a core point, it tries to expand and form a cluster around itself. Based on the experimental results, the authors have demonstrated its robustness toward discovering arbitrarily shaped clusters. 

\subsection{Grid-based Clustering}

In grid-based clustering, the data space is divided into multiple portions (grids) at different granularity levels to be clustered individually. For example, CLIQUE \cite{clique} can automatically find subspaces with high density clusters. No data distribution assumption has been made. 
The empirical results demonstrated that it could scale well with the number of dimensions. Thus it is especially efficient in clustering high-dimensional data.  

\subsection{Correlation Clustering}

Correlation clustering \cite{CC} was motivated from a document clustering problem in which one has a pair-wise similarity function $f$ learned from past data. The goal is to partition the current set of documents in a way that correlates with $f$ as much as possible. In other words, we have a complete graph of N vertices, where each edge is labeled either $+$ or $-$. Our goal is to produce a partition of vertices (a clustering) that agrees with the edge labels. The authors have proved that this problem is a NP-complete problem. Hence they proposed two approximation algorithms to achieve the partitioning.

The first method called \emph{Cautious} is to minimize the disagreements (number of $-$ edges inside clusters plus the number of + edges between clusters), whereas the second method called \emph{PTAS} is to maximize the agreements (number of + edges inside clusters plus the number of $-$ edges between clusters). Basically, the ideas of the above two methods are the same (to aggregate the vertices which agree with their edge labels). The first method is discussed in detail in this work.

First, we arbitrarily choose a vertex v. Then we pick up all the positive neighbors (the neighbor vertices with $+$ edge) of the vertex and put them into a set $A$.  Having picked up all the positive neighbors of the vertex, we perform pruning. That is the 'Vertex Removal Step'.  In this step, we move on to check 3$\delta$-bad for all the positive neighbors of the vertex, where $\delta=1/44$. If there are, we remove it from the set $A$. After the removal step, the next step is 'Vertex Addition Step' in which we try to add back some vertices which are 7$\delta$-good with the chosen vertex v to the set $A$. The vertices in the set $A$ are then chosen as one cluster. The above steps are repeated until no vertices are left or the set $A$ becomes empty.

\subsection{Spectral Clustering}

Some of the existing clustering approaches may find local minima and require an iterative algorithm to find good clusters using different initial cluster starting points. In contrast, spectral clustering \cite{SC,SC2,SC3} is a relatively promising approach for clustering based on the leading eigenvectors of the matrix derived from a distance matrix. The main idea is to make use of the spectrum of the similarity matrix of the data to perform dimensionality reduction for k-means clustering in fewer dimensions. The seminal work \cite{SC} is discussed in this work.

At the beginning, we form an affinity matrix A, which is a NxN matrix and N is the total number of data points. Each entry $A_{ij}$ corresponds to the similarity measure between the data points $s_i$ and $s_j$. The scaling parameter $\sigma^2$ controls how rapidly $A_{ij}$ falls off with the distance between $s_i$ and $s_j$. After we have formed the affinity matrix A, we construct the Laplacian matrix L from the normalized affinity matrix of A.  Then we find the k leading eigenvectors (i.e. with k leading eigenvalues) of L and form the matrix X by stacking the eigenvectors in column. After we have stacked the eigenvectors to form the matrix X, we normalize each row. Then we treat each row in X as a data vector and use k-means clustering algorithm to cluster them. The clustering results are projected back onto the original data (i.e. it assigns the original point $s_i$ to cluster j if and only if row i of the matrix X is assigned to cluster j). 

\subsection{Gravitational Clustering}

Distinct from the works we have mentioned, gravitational clustering is considered as a rather unique method. It was first proposed by Wright \cite{Wright1977151}. In the method, each data instance is considered as a particle within the feature space. A physical model is applied to simulate the movements of the particles. 
As described in  \cite{gc2}, Jonatan et al. proposed a new gravitational clustering method using Newton laws of motion. 
A simplified version of gravitational clustering was proposed by Long et al. \cite{springerlink}.
Wang et al. proposed a local shrinking method to move data toward the medians of their k nearest neighbors \cite{RePEc}. 
Blekas and Lagaris \cite{Blekas:2007:NCA:1231536.1231754} proposed a similar method called Newtonian Clustering in which Newton's equations of motion are applied to shrink and separate data, followed by Gaussian mixture model building.
Molecular dynamics-like mechanism was also applied for clustering by Junlin et al \cite{Junlin20111721}.

\subsection{Herd Clustering}
To tackle the clustering problem, a novel clustering method, Herd Clustering (HC), has been proposed by Wong et al. \cite{wong2014herd}. It novelties lie in two aspects: (1) HC is inspired from the nature, herd behavior, which is a commonly seen phenomenon in the real world including human mobility patterns \cite{peng2012collective}. Thus it is very intuitive and easy to be understood for its good performance. (2) HC also demonstrates that cluster analysis can be done in a non-traditional way by making data \textit{alive}.

HC differs from the traditional ones. Instead of trying hard to analyze data alone, it also spends effort on moving data. Two stages are proposed in HC.

Inspired by the herd behavior \cite{herd}, an attraction model is used to guide data movements in the first stage. Each data instance is represented by a particle. The coordinate position of a particle is given by the values of the corresponding data instance it represents. The particles attract each other if their distances are smaller than a threshold. Each particle has its own velocity (initially zero). In each iteration, the velocity of a particle is affected by the neighborhood particles. If most particles are found in a particular direction, the velocity of the particle is accelerated toward that direction.

After all the iterations in the first stage, all data instances should be well separated and merged together. They are much easier to be clustered than before. Thus an intuitive approach is proposed to cluster data in the second stage. A list of cluster centroids is maintained. At the beginning, the centroid list is empty. For each point, we check whether its distance to any centroid is smaller than the threshold. If a centroid is detected, then the point is assigned the same cluster as the centroid. If its distances to all centroids are higher than or equal to the threshold, the point is added to the list and start a new cluster around it. After all data instances are scanned, a clustering result is obtained.

At the first glance, HC is similar to Gravitation Clustering (GC) \cite{Wright1977151}: data instances are moved according to a model. Nonetheless, their details are totally different. For instance, the model in GC is a physical model following Newton Laws of motion, while that in HC is an artificial model which is designed for computational efficiency. The particle acceleration decreases as the inter-particle distance increases in GC while they are independent in HC. Calculus is involved in GC whereas only computationally efficient operations are allowed in HC.

\subsection{Others}

There are lots of other clustering methods proposed in the past. For instance, Maulik et al. applied a genetic algorithm to search for cluster centers \cite{GA}. A globally incremental approach to k-means has been reported in \cite{globalkmeans}. Celeux et al. have proposed a novel method called Gaussian parsimonious clustering models \cite{gaussian}. Different distance measures have been incorporated into an objective function to cluster arbitrary number of clusters  \cite{CA}. A hierarchical agglomerative clustering methodology using symbolic objects has been described in \cite{symbolic}. Tsao et al. used a fuzzy Kohonen network for clustering \cite{fuzzyKohonen}. A fuzzy c-means algorithm has been developed as described in \cite{Wu20022267,Zhu:2009}. An alternative pruning approach to reduce the noise effect has also been proposed for the fuzzy c-means algorithm \cite{ZhangL03}. In recent years, several kernel methods have been developed for clustering \cite{kernelSurvey}. A fuzzy-rough set application to microarray data has also been reported in \cite{Maji11}. Hu et al. have applied a hierarchical clustering method for active learning \cite{Hu:2009}. Interestingly, Corsini et al. have trained a neural network to define dissimilarity measures which are subsequently used in the relational clustering \cite{1262554}. Gullo et al. have also proposed clustering methods on uncertain data \cite{Gullo:2008:CUD:1434517.1434536,gullo2008hierarchical,gullo2012uncertain}. There are many other works; more details can be found in \cite{clusteringReview,BaraldiB99,Wong:2012:EMO:2181343.2181776}.

\section{Advanced Clustering}

\subsection{Clustering on Data Stream}

The previous clustering methods assume data are static during clustering. Nonetheless, modern data are not static necessarily. In fact, data can be transmitted in streaming form; for instance, real-time financial stock market data, video surveillance data, and social media data. Modern data keeps itself changing and evolving during the course of clustering. For analysis of such data, the ability to process the data in a timely manner with little memory is crucial. In light of that, different data stream clustering methods are proposed. Fo instance, Guha et al. have proposed one of the first-known method, STREAM, to solve the k-median problem on streaming data with constant-factor approximation \cite{Guha:2003:CDS:776752.776777}. An incremental clustering method (COBWEB) has also been proposed to maintain a hierarchical clustering tree on streaming data by Fisher \cite{fisher1995optimization}. Zhang et al. have proposed an efficient data clustering method for large datasets \cite{Zhang:1996:BED:233269.233324}. Thanks to its linear complexity and single-pass nature, it can also be applied to cluster data streams with a tree data structure, CF Tree \cite{Zhang:1996:BED:233269.233324}. On the other hand, an incremental clustering method (C2ICM) has been proposed to data stream clustering problems. In particular, a lower bound for its clustering performance has also been provided \cite{charikar1997incremental}.

\subsection{Clustering on Sequence Data}

In the past years, probabilistic graphical models have been successfully applied to different problems such as gene clustering \cite{pmid16127451,pmid17218491,pmid20445623}. In particular, Hidden Markov Model (HMM) has been demonstrated successful for clustering sequence data in a wide range of domains \cite{Rabiner:1990:THM:108235.108253}. 

\subsubsection{Description}

Hidden Markov Model (HMM) is a probabilistic graphical model which assumes a sequence of symbols is controlled and generated by a corresponding sequence of hidden states with the same sequence length. In particular, Markov property is assumed for the sequence of hidden states; in other words, each hidden state solely depends on its previous hidden state on the same sequence. Although such Markov assumption over-simplifies the independence between different states, it can work fairly well in practice. Moreover, it greatly reduces the computational complexity in HMM learning and inference. Mathematically, an HMM can be described as $\theta$:
\begin{align*}
& \theta=(\{a_{ij}\},\{b_i(x)\},\{\pi_i\}) 
\\
& \forall i,j \in \{1,2,...,N\}, \forall x \in X
\\
& s.t.
\\
& \sum_{i=1}^N \pi_{i} = 1
\\
& \sum_{j=1}^N a_{ij} = 1\ \forall i \in \{1,2,...,N\}
\\
& \sum_{x \in X}  b_i(x) = 1 \ \forall i \in \{1,2,...,N\}
\\
& 0 \leq \pi_{i},a_{ij}, b_i(x) \leq 1 \ \forall i,j \in \{1,2,...,N\}, \forall x \in X
\end{align*}
where $a_{ij}$ is the transition probability from state $i$ to state $j$; $b_i(x)$ is the emission probability to emit $x$ at state $i$; $\pi_i$ is the initial state probability for state $i$.

\begin{figure*}[h!]
\centering
\includegraphics[width=0.7\textwidth]{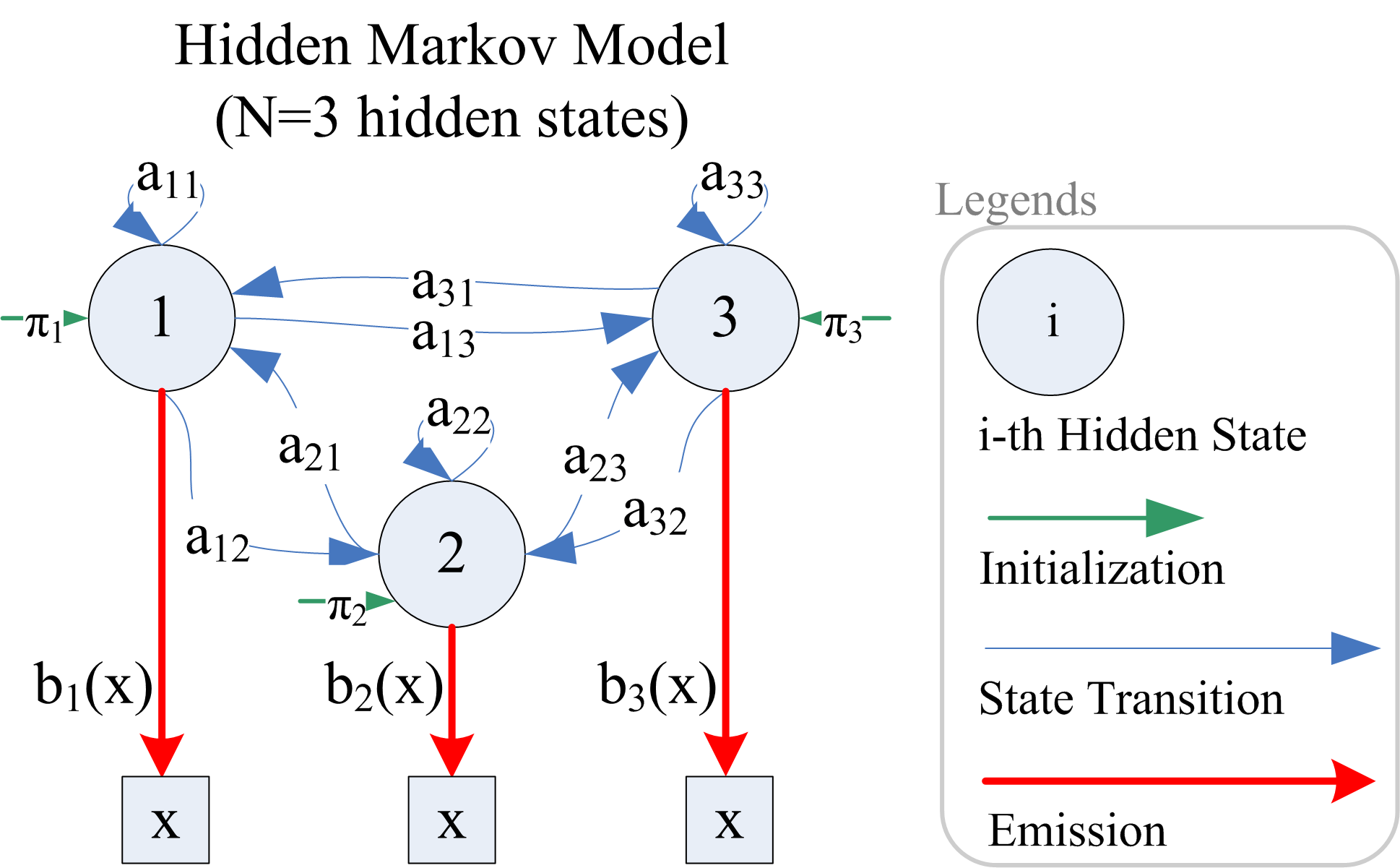}
\caption{Hidden Markov Model (HMM) example with $N=3$ hidden states.}
\label{fig:simpleHMM}
\end{figure*}

For illustrative purposes, an HMM example with $N=3$ hidden states is depicted in Figure \ref{fig:simpleHMM}. In that HMM example, we have 3 hidden states. At the beginning of sequence, we have the initialization probabilities $\{\pi_1,\pi_2,\pi_3\}$ for each hidden state, representing their chances to be the first hidden state. After that, the transition probabilities $\{a_{ij}\}$ determine the next hidden state recursively. For each hidden state traversal, depending on the current state, a symbol $x$ is emitted and appended to form an output sequence based on the emission probabilities $\{b_i(x)\}$.

\subsubsection{Model Learning}

To learn an HMM from sequences, Baum-Welch algorithm is usually applied to learn the unknown parameters \cite{Rabiner:1990:THM:108235.108253}. Mathematically, Baum-Welch algorithm is an Expectation Maximization (EM) algorithm to find the maximal likelihood estimates of the HMM parameters. Thus we would like to note that Baum-Welch algorithm highly depends on the first random initialization iteration and can be trapped in local optima. Multiple runs are usually adopted to circumvent such issues. Mathematically, the Baum-Welch training algorithm can be described herein:
\\
\\
\textbf{Input}: A set of sequences $S=\{s_1,s_2,s_3,...,s_M\}$ of length $L$. Each sequence $s_m$ can be represented as $s_{m} = s_{m1}s_{m2}...s_{mL}$ where $s_{mp} \in X \   \forall m \in \{1,2,...,M\}, \forall p \in \{1,2,...,L\} $.
\\
\\
\textbf{Output:} an HMM model $\theta$ trained  to represent the set of sequences:
\begin{align*}
& \theta=(\{a_{ij}\},\{b_i(x)\},\{\pi_i\}) 
\\
& \forall i,j \in \{1,2,...,N\}, \forall x \in X
\end{align*}
where $a_{ij}$ is the transition probability from state $i$ to state $j$; $b_i(x)$ is the emission probability to emit $x$ at state $i$; $\pi_i$ is the initial state probability for state $i$. 
\\
\\
\textbf{Method:}
At the beginning of Baum-Welch algorithm, we randomly initialize those HMM model parameters $\theta_0$ and iteratively refine them in each iteration. In the expectation step (\textbf{E-step}) of the $l$-th iteration, we calculate the expected values of being in state $i$ based on the current parameter estimates $\theta_l$. Specifically, we calculate:
\begin{align*}
& \gamma_p^m(i) = \frac        {\alpha_p^m(i) \beta_p^m(i)}                {P(s_m;\theta_l)}
\\
& \forall m \in \{1,2,...,M\}, \forall p \in \{1,2,...,L\}, \forall i \in \{1,2,...,N\}
\end{align*}
where $\gamma_p^m(i)$ is the expected probability of being in state $i$ at the $p$-th position of the $m$-th sequence $s_m$; $\alpha_p^m(i)$ and $\beta_p^m(i)$ are the forward and backward probability of the $m$-th sequence $s_m$  to be in state $i$ at the $p$-th position as calculated by the dynamic programming approach \cite{Rabiner:1990:THM:108235.108253}. $P(s_m;\theta_l)$ is the probability of observing $s_m$ given the existing HMM model parameter $\theta_l$ which can be calculated as $P(s_m;\theta_l)=\sum_{i=1}^{N} \alpha_p^m(i) \beta_p^m(i)$. In addition, we also calculate the expected values of state transitions from state $i$ to state $j$:
\begin{align*}
& \zeta_p^m(i,j) = \frac        {\alpha_p^m(i)  a_{ij} b_j(s_{mp})   \beta_{p+1}^m(j)}  {P(s_m;\theta_l)}
\\  
& \forall m \in \{1,2,...,M\},  \forall p \in \{1,2,...,L\},  \forall i,j \in \{1,2,...,N\}
\end{align*}
where $\zeta_p^m(i,j)$ is the expected probability of transiting from state $i$ at the $p$-th position to state $j$ at the ($p+1$)-th position for the $m$-th sequence $s_m$, given the current parameter estimates $\theta_l$.

In the maximization step (\textbf{M-step}) of the $l$-th iteration, those model parameters are refined to be the maximal likelihood estimates for those expected values:
\begin{align*}
\pi_i' &= \frac{\sum_{m=1}^{M} \gamma_1^m(i)}{M} \ \ \forall i \in \{1,2,...,N\} \\
a_{ij}' & = \frac   {\sum_{m=1}^{M} \sum_{p=1}^{L-1} \zeta_p^m(i,j)}    {\sum_{m=1}^{M} \sum_{p=1}^{L-1}  \gamma_p^m(i)}  \ \   \forall i,j \in \{1,2,...,N\} \\
b_{i}(x)' & = \frac   {\sum_{m=1}^{M} \sum_{p=1}^{L}  \gamma_p^m(i) [ s_{mp} = x ]}    {\sum_{m=1}^{M} \sum_{p=1}^{L}  \gamma_p^m(i)} \ \   \\ 
&\forall i \in \{1,2,...,N\},  \forall x \in \{A,C,G,T,-\} \\
\theta_{l+1} & =(\{a_{ij}'\},\{b_i(x)'\},\{\pi_i'\})  \ \  \\
& \forall i,j \in \{1,2,...,N\},  \forall x \in \{A,C,G,T,-\} 
\end{align*}
The new HMM model parameters $\theta_{l+1}$ are used in the next iteration. We repeat the E-step and M-step alternatively until the HMM model parameters are not changed anymore. In other words, the difference between $\theta_{l}$ and $\theta_{l+1}$ converges to a numerically negligible value at which a local optimum is found.

\section{Verification}

\subsection{Benchmark Data Sources}
Benchmark datasets can be downloaded from the UCI Machine Learning Repository \cite{Frank+Asuncion:2010}.

\subsection{Performance Metrics  for Clustering}
For clustering, Rand Index \cite{rand}, Purity \cite{purity}, F-measure \cite{purity}, and Normalized Mutual Information (NMI) \cite{NMI} are usually adopted for performance benchmarking. Rand Index is based on the intra-cluster similarity and inter-cluster dissimilarity. For the intra-cluster similarity, if a pair of data vectors is assigned the same cluster in both the target result and the clustering result, then the score will be increased by one. For the inter-cluster dissimilarity, if a pair of vectors is assigned different clusters in both the target result and the clustering result, then the score will also be increased by one. On the contrary, if a pair of data vectors is in the same cluster in the target result, but not in the clustering result, the score will not be increased. After we have checked all the possible pairs, the score is normalized by the total number of possible pairs. Mathematically, the formula is derived as follows

\[
Rand\ Index = \frac{\sum_{i=1}^{n}\sum_{j=1}^{n}s_{ij}}{n^2-n} \ where \  i\neq j\ ,
\]
\[
   s_{ij} = \left\{ 
  \begin{array}{l l}
    1 &  \mathrm{if }\ G_o(d_i)=G_o(d_j) \  and \   G(d_i)=G(d_j). \\
    1 &  \mathrm{if }\ G_o(d_i)\neq G_o(d_j) \  and \   G(d_i)\neq G(d_j). \\
    0 & \mathrm{otherwise}. \\
  \end{array} \right. 
\]
, where n is the number of data vectors, $d_i$ is the $i$th data vector, $d_j$ is the $j$th data vector, $G_o(d)$ is the cluster group id of a data vector $d$ in the target result, $G(d)$ is the cluster group id of a data vector $d$ in the clustering result. On the other hand, F-measure is similar to Rand Index with the exception that true negatives are not taken into account.  Mathematically, the formula is derived as follows:
\[
F-measure = \frac{2a}{2a+b+c}
\]
\[
  \begin{array}{lll}
    a & = & \sum_{i=1}^{n}\sum_{j=1}^{n} [i \neq j][G_o(d_i)=G_o(d_j) \  \&  \   G(d_i)=G(d_j)]. \\
    b & = & \sum_{i=1}^{n}\sum_{j=1}^{n} [i \neq j][G_o(d_i) \neq G_o(d_j) \  \&  \   G(d_i)=G(d_j)]. \\
    c & = & \sum_{i=1}^{n}\sum_{j=1}^{n} [i \neq j][G_o(d_i)=G_o(d_j) \  \&  \   G(d_i) \neq G(d_j)]. \\
  \end{array}
\]
, where [...] is the Iverson bracket. In contrast, purity solely measures the intra-cluster similarity. Nevertheless, it is useful in the sense that we only care about the quality of individual clusters. Mathematically, the purity of a cluster $C_i$ of size $n_i$ is defined below. For $n$ data instances with $k$ cluster groups, the overall purity of a clustering result is defined as:
\[
P(C_i)=\frac{1}{n_i}\max_j (n^j_i)
\]
\[
Purity=\sum^k_{i=1} \frac{n_i}{n} P(C_i)
\]
, where $n^j_i$ is the number of the data instances of $j$th class that are assigned to the $i$th cluster. To account for all the performance results, Normalized Mutual Information (NMI) can also be used \cite{NMI}. For all non-deterministic methods, the performance metrics are taken by averaging over multiple runs. For all deterministic methods, the performance metrics are taken by running once only.

\subsection{Performance Metrics for Prediction}
From the perspective of predictive tasks, a clustering outcome can be categorized into 4 types. If the clustering outcome is consistent with the truth, it is called either \textbf{True Positive (TP)} or \textbf{True Negative (TN)}, depending on the actual value. Otherwise, it is called \textbf{False Positive (FP)} or \textbf{False Negative (FN)} respectively. In different problem domains, FPs and TNs are depreciated and weighted differently. For instance, FPs are more tolerated than FNs in human disease diagnosis.
%

To summarize the prediction performance of a clustering method, accuracy is widely adopted. It is defined as follows:
\[
Accuracy = \frac{TPs+NPs}{TPs+FNs+FPs+TNs}
\]
Nonetheless, accuracy may be non-informative if the dataset is imbalanced or mis-clustering cost is very high. For instance, if only the performance of a method on positive class prediction is practically interesting, we can adopt precision and sensitivity (a.k.a. true positive rate and recall) which are defined as follows:
\[
Precision = \frac{TPs}{TPs+FPs}
\]
\[
Sensitivity = \frac{TPs}{TPs+FNs}
\]
Alternatively, F-measure can be applied to combine precision and sensitivity into a single performance metric. It is defined as the harmonic mean of precision and sensitivity. The duals of precision and sensitivity for negative class clustering are negative predictive value (NPV) and specificity respectively.
\[
NPV = \frac{TNs}{TNs+FNs}
\]
\[
Specificity = \frac{TNs}{TNs+FPs}
\]
In particular, we would like to note that the well-known false positive rate (FPR) and false discovery rate (FDR) are defined as follows:
\[
FPR = 1 - Specificity\ , \ FDR = 1 - Precision
\]
Although the performance metrics described are very suitable for evaluating discrete clustering predictions. Nonetheless, the modern clustering methods usually assign a confidence value to each of its prediction. To examine the modern methods in full spectrum, receiver operating characteristics (ROC) curves and precision-recall (PRC) curves are proposed. Different thresholds are cut at the confidence values to observe the performance trade-off of each method. For instance, the trade-off between sensitivity and false positive rates can be observed from ROC curves whereas that between precision and recall can be observed from PRC curves. The area under ROC curves (AUC) is usually adopted as a benchmarking metric.

\subsection{Evaluation Procedures}
The most typical evaluation procedure is to divide a dataset into two sets: training dataset and testing dataset. The training dataset is used for training a clustering model, while the testing dataset is isolated and reserved for testing the trained model. In particular, the most common procedure is N-fold cross-validation which has $N$ iterations. The dataset is randomly divided into N non-overlapping subsets. In each iteration, a subset is rotated as the testing dataset while the others are assigned as the corresponding training dataset. If the input data is scarce or costly, leave-one-out cross-validation can also be applied. In that case, only one data sample is left out for testing, while the others are allocated as the training dataset  in each iteration. 

\subsection{Statistical Tests}
Since some of the existing clustering methods are stochastic, multiple replicate runs need to be executed for comprehensive benchmarking \cite{wong2014herd}. The means and standard deviations of performance metrics are usually reported for fair comparison. To justify the results, statistical tests are adopted to assess the statistical significances; For instance, t-tests, Mann-Whitney U-tests (MWU), and Kolmogorov-Smirnov test (KS).

\section{Benchmarking}
\begin{figure*}[hp!]
	\centering
	\includegraphics[width=\textwidth]{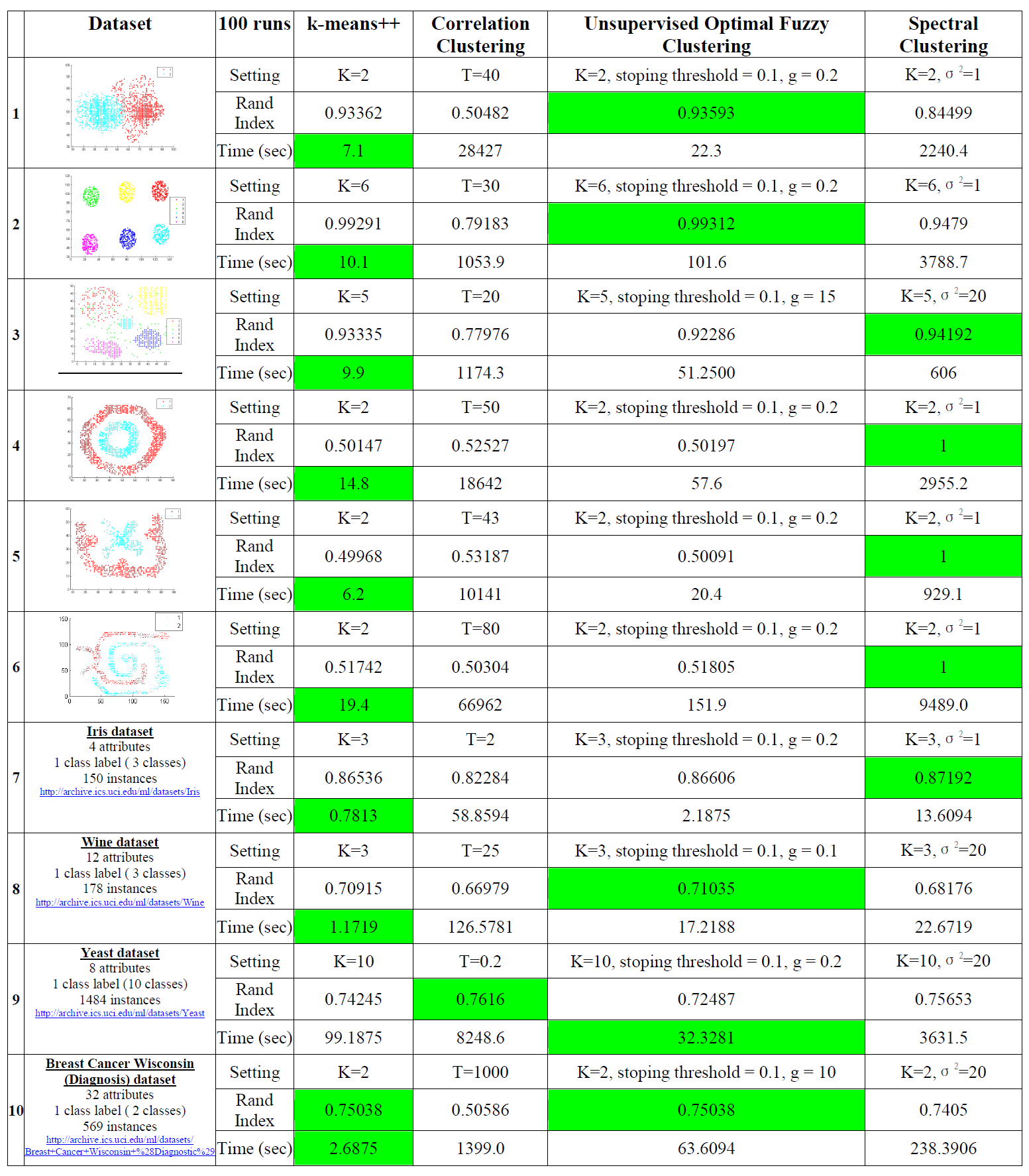}
	\caption{Performance Comparison between the methods selected.}
	\label{fig:benchmarking}
\end{figure*}

To investigate the performance difference between those methods, four representative methods have been selected and run on different datasets. K-means++ is chosen for its simplicty and superior performance over the traditional k-means method; Correlation clustering is selected to represent the algorithms with solid theoretical support; Unsupervised optimal fuzzy clustering is chosen to represent the soft clustering algorithms; Spectral clustering is selected to represent the modern clustering algorithms. Since all of the methods selected are stochastic, 100 replicate runs are executed to compute the average performance metrics for each method on each dataset. All the parameters were tuned for each algorithm and dataset manually. The results are depicted in Fig. \ref{fig:benchmarking}.

From the results, we can observe that the clustering methods exhibit different characteristics on different datasets. In general, based on the performance metric (Rand Index), spectral clustering is found to perform the best among the selected algorithms whereas the performance of correlation clustering is relatively limited. Based on the time taken, k-means++ is the fastest one, whereas correlation clustering is the slowest one on most datasets.
The top three datasets are the most typical datasets. Each cluster forms a globular shape. It is not hard for us to expect that they can be solved by most clustering algorithms. The result turns out to concede with our expectation, except correlation clustering. 
The middle three datasets are difficult datasets. Each cluster is an irregular shape. Within the same dataset, each cluster is even not guaranteed to be similar to the other clusters. Interestingly, a nearly perfect result can be obtained by spectral clustering, reflecting that the dimensional transformation ability within spectral clustering does play a role in lowering the difficulties in handling such irregular data shapes.
The bottom four datasets are the well-known datasets taken from the UCI machine learning repository. The number of attributes is ranged from 4 to 32. The number of class labels is ranged from 2 to 10. The number of instances is ranged from 150 to 1484. In the experiment, each algorithm has managed to perform well on a particular dataset. No conclusive insights can be drawn from the result. The data dependency of the clustering algorithms is fully reflected on those datasets.

\section{Summary and Future Works}

\subsection{Summary}

With growing data, cluster algorithms (also known as cluster analysis) become important tools for analyzing data.  In this book chapter, we have reviewed the existing clustering algorithms from different paradigms: Partitional Clustering, Hierarchical Clustering, Density-based Clustering, Grid-based Clustering, Correlation Clustering, Spectral Clustering, Gravitational Clustering, Herd Clustering, and Others. Especially, we have focused on their methodologies and design concepts. Advanced clustering methods have also been reviewed; for instance, data stream clustering and sequence clustering. 

To verify the algorithms' competitiveness,  different types of performance metrics have been defined and reviewed. In particular, benchmark studies have been conducted to observe the empirical performance of the selected methods: k-means++, correlation clustering, fuzzy clustering, and spectral clustering. The numerical results reveal that spectral clustering has its own competitive edge over the other methods on low-dimensional datasets. For high-dimensional datasets, we cannot observe any significant performance difference between the selected methods. 

Nonetheless, during the course of the studies here, we found several future directions which we believe they are promising. They are described in the following section.

\subsection{Future Works}

\subsubsection{Computational Scalability}
As mentioned at the very beginning of this book chapter, the recent advancements of science and technologies enable massive data generation in recent years. Some of the existing computational methods may not scale with the large amount of data. For instance, the high computational complexity of spectral clustering method \cite{pmid7584439} is no longer practical to be run on the current datasets. It is imperative for us to develop new and scalable methods to keep in pace with the data generation speed.

\subsubsection{Advanced Learning Methods}
In this book chapter, we have provided an overview on clustering. It is undeniable that other machine learning methods can be applied as well \cite{wong2014signalspider}; for instance, probabilistic graphical models can be developed and applied to capture/eliminate the uncertainty and noises in real world data.

\subsubsection{Domain Knowledge}
The existing clustering algorithms are built for general purposes. Domain knowledge can be incorporated if a clustering algorithm is applied to a specific task; for instance, if data is sparse, a sparse clustering algorithm can be applied to boost up the execution speed.

\bibliographystyle{IEEEtran}
 \bibliography{BCclustering}

\end{document}